\documentstyle[12pt]{article}

\newlength{\dinwidth}
\newlength{\dinmargin}
\begin{document}

\def\bold#1{\setbox0=\hbox{$#1$}%
     \kern-.025em\copy0\kern-\wd0
     \kern.05em\copy0\kern-\wd0
     \kern-.025em\raise.0433em\box0 }
\def\slash#1{\setbox0=\hbox{$#1$}#1\hskip-\wd0\dimen0=5pt\advance
       \dimen0 by-\ht0\advance\dimen0 by\dp0\lower0.5\dimen0\hbox
         to\wd0{\hss\sl/\/\hss}}
\def\lq{\left [}
\def\rq{\right ]}
\def\LL{{\cal L}}
\def\VV{{\cal V}}
\def\AA{{\cal A}}
\def\MM{{\cal M}}

\newcommand{\be}{\begin{equation}}
\newcommand{\ee}{\end{equation}}
\newcommand{\bea}{\begin{eqnarray}}
\newcommand{\eea}{\end{eqnarray}}
\newcommand{\nn}{\nonumber}
\newcommand{\dd}{\displaystyle}
\newcommand{\bra}[1]{\left\langle #1 \right|}
\newcommand{\ket}[1]{\left| #1 \right\rangle}
\thispagestyle{empty}
\vspace*{1cm}
\rightline{BARI-TH/93-144}
\rightline{June 1993}
\vspace*{2cm}
\begin{center}
  \begin{Large}
  \begin{bf}
Radiative heavy meson transitions \\
  \end{bf}
  \end{Large}
  \vspace{8mm}
  \begin{large}
P. Colangelo $^{a,}$ \footnote{e-mail address:
COLANGELO@BARI.INFN.IT},  F. De Fazio $^b$, G. Nardulli $^{a,b}$\\
  \end{large}
  \vspace{6mm}
$^{a}$ Istituto Nazionale di Fisica Nucleare, Sezione di Bari, Italy\\
  \vspace{2mm}
$^{b}$ Dipartimento di Fisica, Universit\'a
di Bari, Italy \\

\end{center}
\begin{quotation}
\vspace*{1.5cm}
\begin{center}
  \begin{bf}
  ABSTRACT
  \end{bf}
\end{center}
\vspace*{0.5cm}

We evaluate the  radiative and hadronic decay rates of the
$D^*$ mesons using the Heavy Quark Effective Theory and the Vector Meson
Dominance hypothesis.  We also estimate the width of the $B^*$
electromagnetic transitions and
the radiative decays of positive parity $J^P=0^+, 1^+$ charmed
mesons.
\noindent
\end{quotation}

\newpage
\baselineskip=18pt
\setcounter{page}{1}

In this letter we wish to show that the heavy quark and chiral $SU(3)_L \times
SU(3)_R$ symmetries of QCD, together with the Vector Meson Dominance (VMD)
hypothesis, can be used to relate radiative and semileptonic charmed meson
decays. We shall show that, using experimental data on the
$D \to \pi \ell \nu_\ell$, $D \to K \ell \nu_\ell$
 and $D \to K^* \ell \nu_\ell$ transitions as an input, it
is possible to predict  the radiative branching ratios $D^* \to D \gamma$,
the hadronic rate $D^* \to D \pi$ and therefore the full $D^*$ width. Moreover,
the heavy quark flavour symmetry will provide us with a prediction of the $B^*$
decay rate. We shall also employ information on the positive parity
$J^P=0^+ ,\;  1^+$ charmed mesons (masses and couplings) from the analysis of
semileptonic $D$ decays via axial currents in order to get predictions on the
radiative decays of these states as well.

The calculation is based on a chiral lagrangian approach
to the heavy and light meson interactions incorporating the chiral symmetry and
the heavy quark spin flavour symmetry
\cite{WISE, CASAL}.
At the lowest order in the light meson derivatives, the chiral
lagrangian can be written as follows:

\be \LL=\LL_0 + \LL_2 + ... \hskip 3 pt , \label{eq : 1}\ee

\noindent where $\LL_0$ contains the light meson matrix $\Sigma$ and the
heavy meson fields $H$:

\be
\LL_{0}=\frac{f_{\pi}^2}{8}<\partial^\mu\Sigma\partial_\mu
\Sigma^\dagger > +i < H_b v^\mu D_{\mu ba} {\bar H}_a >
+i g <H_b \gamma_\mu \gamma_5 \AA^\mu_{ba} {\bar H}_a>
\hskip 3 pt ,\label{eq : 2}
\ee

\noindent whereas $\LL_2$ \cite{CASAL}
describes the interactions with the light
$J^P=1^-$ mesons:

\bea
\LL_2&=& -\frac{f^2_{\pi}}{2}a <(\VV_\mu-
\rho_\mu)^2>+\frac{1}{2g_V^2}<F_{\mu\nu}(\rho)F^{\mu\nu}(\rho)> \nn\\
&+&i\beta <H_bv^\mu\left(\VV_\mu-\rho_\mu\right)_{ba}{\bar H}_a>\nn\\
&+&\frac{\beta^2}{2f^2_{\pi} a}<{\bar H}_b H_a{\bar H}_a H_b>+
i \lambda <H_b \sigma^{\mu\nu} F_{\mu\nu}(\rho)_{ba} {\bar H}_a>
\hskip 3 pt . \label{eq : 3}\eea

\noindent In eqs.(\ref{eq : 2},\ref{eq : 3})
 $<\ldots >$ means the trace, $f_{\pi}=132 \hskip 3 pt MeV$ and:

\bea
D_{\mu ba}&=&\delta_{ba}\partial_\mu+\VV_{\mu ba}
=\delta_{ba}\partial_\mu+\frac{1}{2}\left(\xi^\dagger\partial_\mu \xi
+\xi\partial_\mu \xi^\dagger\right)_{ba} \hskip 3 pt ,\label{eq : 4}\\
\AA_{\mu ba}&=&\frac{1}{2}\left(\xi^\dagger\partial_\mu \xi-\xi
\partial_\mu \xi^\dagger\right)_{ba} \hskip 5 pt ; \label{eq : 5}
\eea

\noindent the field  $\xi$ is defined by:

\be
\xi=\sqrt{\Sigma}=e^{iM/f_{\pi}} \label{eq : 6} \ee

\noindent
where $M_{ba}$ is the usual $3 \times 3$ matrix describing the octet of pseudo
Nambu-Goldstone bosons.
The $0^-(P)$ and $1^-(P^*)$ $Q {\bar q}_a$ heavy mesons are described by the
effective fields:

\bea
H_a &=& \frac{(1+\slash v)}{2}[P_{a\mu}^*\gamma^\mu-P_a\gamma_5] \hskip 3 pt ,
\label{eq : 7}\\
{\bar H}_a &=& \gamma_0 H_a^\dagger\gamma_0 \hskip 3 pt , \label{eq : 8}
\eea
\noindent where $v$ is the heavy meson four-velocity, $a=1,2,3$
(for $u,d$ and $s$ respectively), and
$P^{*\mu}_a$ and $P_a$ are annihilation operators normalized as follows:

\bea
\langle 0|P_a| Q{\bar q}_a (0^-)\rangle & =&\sqrt{m_P};\nonumber \\
\langle 0|P^*_a| Q{\bar q}_a (1^-)\rangle & = & \epsilon^{\mu}\sqrt{m_P^*}
\nonumber \eea

\noindent (with $m_P=m_{P^*}$ in the limit of heavy quark spin symmetry).\par
The coupling of the light vector meson resonances belonging to the $1^-$ low
lying nonet (with $\phi=s {\bar s}$) has been introduced in $\LL_2$
(eq.(\ref{eq : 3})) using
the hidden gauge symmetry approach \cite {BANDO}; $\rho^{\mu}$ contains the
light $1^-$ fields, normalized according to :

\be \rho^{\mu}=i \; {g_V \over \sqrt{2} } \hat{\rho}^{\mu} \label{eq : 9} \ee

\noindent where $\hat{\rho}$ is the $3 \times 3$ matrix describing the $1^-$
light meson nonet; $F_{\mu \nu}(\rho)=
\partial_{\mu} \rho_{\nu} - \partial_{\nu}
\rho_{\mu} + [\rho_{\mu}, \rho_{\nu}]$ and $g_V$, $a$ and $g$ are coupling
constants. The two KSRF relations \cite{KSRF} fix the value of $a$ and $g_V$:

\be a=2 \hskip 2 cm g_V=5.8 \hskip 6 pt . \label{eq : 10} \ee

The Lagrange density (\ref{eq : 1}) is the most
general expression displaying chiral invariance in the lowest order in the
meson derivatives; this invariance can be proved observing that under
$SU(3)_L \times SU(3)_R$ transformations:

\bea \xi \to L \xi U^{\dag}= U \xi R^{\dag}  \hskip 3 pt , \nonumber \\
 \AA^{\mu} \to U \AA^{\mu} U^{\dag} \hskip 3 pt , \label{eq : 11} \\
 D^{\mu} {\bar H} \to U D^{\mu} {\bar H} \hskip 3 pt . \nonumber \eea

\noindent The Lagrangian (\ref{eq : 1})
displays also the heavy quark flavour symmetry (since the
mass of the heavy quark does not appear in $\LL$) and the heavy quark spin
symmetry, because, under the heavy quark spin group $SU(2)_v$ the fields
transform as follows:

\be H_a \to S H_a \hskip 3 pt; \hskip 2 cm
{\bar H}_a \to {\bar H}_a S^{\dag} \hskip 3 pt , \label{eq : 12} \ee

\noindent with $SS^{\dag}=1$ and $[\slash v, S]=0$.\par
Let us now discuss the constants $g$ and $\lambda$
($\beta$ will not be used hereafter);
$g$ is responsible for the
$D^* D \pi$ coupling and is related to the $D^*$ hadronic width by the
tree level formula:

\be \Gamma(D^{*+} \to D^0 \pi^+)={g^2 \over 6 \pi f_{\pi}^2} |{\vec p}_\pi|^3
\hskip 3 pt ; \label{eq : 13} \ee

\noindent
($\Gamma(D^{*+} \to D^+ \pi^0)$ is smaller by a factor of 2).
As shown in \cite{CASAL}, the coupling  $g$ can be obtained
by the $D \to \pi \ell \nu_{\ell}$ or
 $D \to K \ell \nu_{\ell}$ decay processes assuming a polar $t$-dependence
of the
vector form factor as well as a  determination of the
weak current of the chiral theory \cite{WISE} based on QCD sum rules
\cite{PIETRO}:

\be L^{\mu}=i {\alpha \over 2} <\gamma^\mu (1- \gamma_5) H_b \xi^{\dag}_{ba}>
\hskip 3 pt . \label{eq : 14} \ee

\noindent The parameter
$\alpha$ is related to the leptonic constant $f_P$ defined by:

\be <0|\bar{q}^a \gamma^\mu \gamma_5 Q|P_a(p)>=i f_P p^\mu \hskip 3 pt ,
\label{eq : 15} \ee

\noindent by the relation $\alpha=f_P \sqrt{m_P}$ (neglecting logarithmic
corrections). Since $1/m_Q$ corrections appear to be significant in the
$f_B/f_D$ ratio, one could choose to include them and determine $g$ from the $D
\to \pi, K \ell \nu_\ell$ decays by using for $f_D$ the value $\simeq 200
\hskip 3 pt MeV$, indicated by both QCD sum rules \cite{PIETRO} and lattice
calculations \cite{ABADA}. However, as shown in \cite{CASAL2}, this procedure
produces weak hadronic matrix elements which, when used to predict non leptonic
$B$ decays in the factorization approximation, give results in disagreement
with data. Therefore,
following \cite{CASAL2} we choose to fix $\alpha$ from the value of $f_B$
obtained by QCD calculations \cite{PIETRO, NEUBERT}
which, given the actual value of
$m_b$, much larger than $m_c$, should provide a better approximation.
In \cite{NEUBERT} the QCD sum rules method is applied to the evaluation of
$f_B$ in the infinite heavy quark mass limit, with the result:

\be \alpha=0.35 \; - \; 0.45 \hskip 3 pt GeV^{3 \over 2} \;.
\label{eq : alpha} \ee

\noindent Theoretical uncertainties of this result are significant because the
$O(\alpha_s)$ corrections are more than $ 50 \%$ of the result. Nevertheless
the result (\ref{eq : alpha}) is confirmed by another \cite {PIETRO} QCD sum
rules calculation of $f_B$, performed for finite value of $m_b$, which gives
$\alpha= f_B \sqrt{ M_B} \simeq 0.47 \pm 0.04 \hskip 3 pt GeV^{3/2}$.
Using (\ref{eq : alpha}) and semileptonic
$D$ decays data ($D \to \pi \ell \nu$ and $D \to K \ell \nu$)
we get:

\be |g|=0.40 \pm 0.08 \hskip 3 pt , \label{eq : 16} \ee

\noindent where the error is only theoretical and is a consequence of the
uncertainty in eq.(\ref{eq : alpha});
to this theoretical error one should add a
further $20 \%$ experimental uncertainty from
$D \to \pi (K) \ell \nu$ decay data which give
$|\lambda \alpha| = 0.16 \; GeV^{1 \over 2}$.
Using the same value for $\alpha$, from $D \to K^* \ell \nu_\ell$
decay data \cite{STONE} one obtains for $\lambda$:

\be |\lambda|=0.40 \pm 0.08 \; GeV^{-1} \hskip 3 pt , \label{eq : 17} \ee

\noindent where again the error is only theoretical
(the experimental error is a further $20 \%$).\par
Let us now turn to the decay:

\be D^* \to D \gamma \hskip 3 pt , \label{eq : 18} \ee

\noindent whose matrix element can be written as follows:

\be \MM(D^* \to D \gamma) =e \; \epsilon^{* \mu} J_\mu \hskip 3 pt
\label{eq : 19} \ee

\noindent with:

\bea J_\mu &=& <D(p^{\prime})|J_\mu^{em}|D^*(p, \eta)>=
\nonumber \\
&=& <D(p^{\prime})|e_Q \bar{Q} \gamma^\mu Q + e_q \bar{q} \gamma^\mu q|D^*(
p, \eta)>= \label{eq : 20} \\
&=& e_Q J_\mu^Q + e_q J_\mu^q \nonumber \hskip 3 pt , \eea

\noindent where $e_Q={2 \over 3}$ is the heavy quark ($Q=c$) charge and $e_q$
is the light quark charge ($e_q=e_u=2/3$ for $D^{*0}$ and $e_q=-1/3$ for
$D^{*+}$ and $D^*_s$). Let us consider the two currents appearing in
(\ref{eq : 20})
separately. $J_\mu^Q$ can be expressed in terms of the Isgur-Wise
universal form factor \cite{WI} as follows:

\be <D(p^{\prime})|\bar{c} \gamma^\mu c|D^*(p, \eta)>=
    i  \sqrt{m_D m_{D^*}} \xi(v \cdot v^{\prime})
\epsilon^{\mu \nu \alpha \beta}
\eta_\nu v_\alpha v^{\prime}_\beta  \hskip 3 pt , \label{eq : 21} \ee

\noindent where $p^{\prime}=m_D v^{\prime}$, $p=m_{D^*} v$ and $v
\cdot v^{\prime} \simeq 1$ because:

\be 0=q^2=m^2_D + m_{D^*}^2 -2 m_D m_{D^*} v \cdot v^{\prime} \hskip 3 pt .
\label{eq : 22} \ee

As to the computation of the current

\be J^q_\mu=<D(p^{\prime})|\bar{q} \gamma^\mu q|D^*(p, \eta)> \hskip 3 pt ,
\label{eq : 24} \ee

\noindent we assume VMD hypothesis and
write:

\be J^q_\mu= \sum_{V, \lambda}  <D(p^{\prime}) V(q, {\epsilon_1}(\lambda))|
D^*(p, \eta)> { i \over  q^2-m_V^2}
<0|\bar{q} \gamma_\mu q|V(q, {\epsilon_1}(\lambda))>
 \label{eq : 25} \ee

\noindent where $q^2=0$ and the sum is over the vector meson resonances $V=
\omega
$, $\rho^0$, $\phi$ and over the $V$ helicities.
The vacuum-to-meson current matrix
element appearing in (\ref{eq : 25})
is given, assuming $SU(3)$ flavour symmetry, by:

\be <0|\bar{q} T^i \gamma^\mu q|V(q, { \epsilon_1})>
={\epsilon_1}^\mu f_V \;
Tr(V T^i) \hskip 3 pt , \label{eq : 26} \ee

\noindent where  $(T^i)_{mn}=\delta_{im} \delta_{in}$ and $i=1,2,3$ for $u,d,s$
respectively. From $\omega \to e^+ e^-$ and $\rho^0 \to e^+ e^-$ decays
\cite{PDG} we get the same value for $f_V$: $f_V=0.17 \hskip 3 pt GeV^2$; from
$\phi \to e^+ e^-$ we have $f_{\phi}=f_V + \delta f$ with $\delta f=0.08
GeV^2$, showing a significant $SU(3)$ violation. Using (\ref{eq : 26}) and the
strong lagrangian $\LL_2$ we can easily compute $J^q_\mu$ and therefore
(\ref{eq : 20}). The results are :

\be \MM(D^* \to D \gamma)=i \; \epsilon^{\mu \nu \alpha \beta}
\epsilon^*_\mu \eta_\nu v_\alpha v^{\prime}_\beta \sqrt{m_D m_{D^*}}[e_Q-e_q 2
\sqrt{2} g_V \lambda m_{D^*} {f_V \over m^2_{\omega}}] \hskip 3 pt ,
\label{eq : 27} \ee

\be \MM(D^*_s \to D_s \gamma)=i \; \epsilon^{\mu \nu \alpha \beta}
\epsilon^*_\mu \eta_\nu v_\alpha v^{\prime}_\beta \sqrt{m_{D_s} m_{D^*_s}}
[e_Q+{1 \over 3} 2 \sqrt{2} g_V \lambda m_{D^*_s} {f_{\phi} \over m^2_{\phi}}]
\hskip 3 pt ,\label{eq : 28} \ee

\noindent where $e_Q=e_c={2 \over 3}$.
Eq.(\ref{eq : 27}) holds for both $D^{*+} \to
D^+ \gamma$ and  $D^{*0} \to D^0 \gamma$ (with $e_q=-{1 \over 3}$ and ${2 \over
3}$ respectively), assuming $m^2_\rho \simeq m^2_\omega$. Since (\ref{eq : 17})
only gives the absolute value of $\lambda$,
we have fixed its sign by imposing that
the relative sign between the two contributions is identical to the one given
by the constituent quark model \cite{EICHTEN}, i.e. we take $\lambda=-0.40 \pm
0.08$.
It is worth observing that eqs.(\ref{eq : 27},\ref{eq : 28}) describe with
obvious changes also $B^*$ radiative decays.
\par
{}From the amplitudes (\ref{eq : 27},\ref{eq : 28}) and from eq.(\ref{eq : 13})
we can compute decay rates and branching ratios (BR) for $D^*$ and
$B^*$ decays. They are reported in Table I together with the CLEO data
\cite{CLEO} on  radiative $D^*$ decays.
We observe an overall
agreement between theoretical results and experiment. In particular,
the tiny decay rate $D^{*+} \to D^+ \gamma$ can be explained as an effect of a
cancellation between the two contributions appearing in (\ref{eq : 27}). \par
Let us compare our results with previous work. The general
structure of the matrix elements (\ref{eq : 27}) and (\ref{eq : 28})
 coincides with
previous analyses \cite{EICHTEN, ELETSKY, AMUNDSON, GEORGI, CHENG};
the main differences are in the determination of the light quark current that
is not provided by the heavy quark effective theory. Our outcome coincides with
the constituent quark model result: $\MM(D^* \to D \gamma) \sim ({e_Q \over
m_Q} + {e_q \over m_q})$ with $(m_q)^{-1}=-2 \sqrt{2} \lambda g_V {f_V \over
m_V^2}$ (see \cite{EICHTEN, CHENG}). Whereas $m_q$ in the quark model
coincides with the constituent light quark mass (e.g. $m_u \simeq m_d \simeq
300 \hskip 3 pt MeV$) \cite{EICHTEN}, in our case the mass parameter $(-2
\sqrt{2} \lambda g_V {f_V / m_V^2})^{-1}$ has a significantly larger value,
i.e. $0.55 \hskip 3 pt GeV$ and $0.95 \hskip 3 pt GeV$ for $D^*$ and $D^*_s$
decays respectively.\par
The input value for $g$ (eq. (\ref{eq : 16})
and the result for $(m_q)^{-1}$ are
compatible with the analyses of ref. \cite{AMUNDSON} where $g$ and $m_q$ are
treated as free parameters; in that paper also $SU(3)$
breaking through one loop diagrams are reported. The breaking of $SU(3)$
given by (\ref{eq : 28}) is actually different from that one,
since it is due to explicit $SU(3)$ violation (by $m_\phi \neq m_\omega$
and $f_\phi \neq f_\omega$) and not to chiral loop effects. \par
Let us now turn to the radiative decays of positive parity charmed meson
resonances. As well known \cite{IW, LUKE} in HQET one expects four states:
two of these states,
the $J_P=0^+$ ($D_0$)
and  $J_P=1^+$ ($D'_1$)  form a mass doublet and are
 characterized by a total angular momentum of the
light degrees of freedom $s_\ell= {1 \over 2}$;  a
mass doublet is also formed by the two states having
$s_\ell= {3 \over 2}$ with  $J_P=1^+$ ($D_1$)
and $J_P=2^+$ ($D_2$) \hskip 5pt
\footnote{We shall neglect the possible mixing between $D'_1$ and $D_1$ states
\cite{LUKE,Korner} }.
These states are described by effective operators
analogous to the fields $H_a$ introduced in eq.(\ref{eq : 7}); in particular,
the states having $s_\ell= {1 \over 2}$ are described by:

\be S_a = {1 +  {\slash v} \over 2} [D^{\prime \mu}_1 \gamma_\mu \gamma_5 -
D_0 ] \hskip 5pt; \label{eq : s} \ee

\noindent for these states we write the effective
couplings to the light vector and heavy
negative parity mesons as follows \cite{CASAL}:
\be
\LL^{\prime}=i \zeta <{\bar S}_a H_b \gamma_\mu (\VV^\mu -\rho^\mu)_{ba}>+
i \mu <{\bar S}_a H_b \sigma^{\lambda \nu} F_{\lambda \nu}(\rho)_{ba}> + h.c.
\label{eq : 29}  \ee

The analysis of the $D \to K^*$ semileptonic decays via axial current,
using a polar dependence of the form factors
\footnote{The $t$-dependence of the form factor which describes the
semileptonic $D \to K \ell \nu$ decay has been measured in \cite{nmr}
and found compatible with a polar behaviour. For the form factors
describing $D \to K^* \ell \nu$, no experimental information is available yet.
{}From the theoretical point of view, lattice results
are compatible with a polar dependence \cite{nmr1} whereas QCD Sum Rules
seem  to exclude it \cite{nmr2} for axial currents.},
provides the result \cite{CASAL}:
\be \mu= -0.13 \pm 0.05 \; GeV^{-1} \label{eq : 30} \ee

\par
Similarly to the $D^*$ radiative decays, in order to describe the radiative
processes
\bea
D_0 & \to & D^* \gamma \nonumber \\
D'_1 & \to & D \gamma \nonumber \\
D'_1 & \to & D^* \gamma  \label{eq : 31} \eea
\par
\noindent we distinguish in the
electromagnetic matrix element the term containing the heavy quark current
$ e_c \; \bar c \gamma_\mu c$ and the term with the light quarks
$ e_q \; \bar q \gamma_\mu q$. The first can be obtained from the Lagrange
density:
:

\be \LL^{\prime \prime}=-{e \over 2 m_Q} e_Q {\bar h}_v \sigma^{\mu \nu} h_v
F_{\mu \nu} \hskip 5 pt , \label{eq : lag} \ee

\noindent which allows the transition $Q \to Q \gamma$
\footnote{The use of (\ref{eq : lag}) for $D^* \to D \gamma$ produces exactly
the same result already obtained in the leading $1 \over m_Q$ expansion.}
; the matrix elements
involve the universal form factor
$\tau_{1/2}$, analogous to the Isgur-Wise function,
which has been introduced in \cite{IW} and computed
in \cite{CNP} by QCD sum rules.
On the other hand, the matrix element of the light quark current $e_q {\bar q}
\gamma_\mu q$ can be related, via VMD, to the
coupling in eq.(\ref{eq : 29}). However, in this case the contribution
of the $\zeta$ term, used together with the VMD hypothesis, displays
a breaking of gauge invariance due to non-leading $1 \over m_Q$ terms.
In principle, gauge invariance could be recovered
adding further couplings, but this procedure would spoil
the predictivity of the method. Since no
experimental information is available on  $\zeta$, we choose
$\zeta=0$ in our estimate of the positive parity radiative transitions.
This should give at least an order of magnitude estimate of the $0^+$, $1^+$
radiative decay widths.
The obtained results are given in Table II.
We observe that the radiative widths of positive resonances are small
due to an almost complete cancellation between the two contributions of the
e.m. current.
We also note that for the neutral charged resonances, the values in
Table II would correspond to branching ratios of the order $10^{-4} - 10^{-3}$
\cite{LUKE,Korner}.
\par
In conclusion, we have shown that chiral heavy meson theory can be used to
relate semileptonic $B$ and $D$ decays to charmed resonances decays
by using the additional
hypothesis of Vector Meson Dominance for the light quark vector current. Even
though our results have potentially sizable uncertainties ($1/m_Q$ corrections
to HQET, correction to VMD, etc.), by  using semileptonic decays as an input
we have found results that are in fairly good
agreement with the $D^*$ decay rates; moreover, heavy flavour and chiral
symmetries  have been
used to predict the decay width of $D^*_s$ and $B^*$ mesons.

\newpage
\begin{center}
  \begin{Large}
  \begin{bf}
  Tables Captions
  \end{bf}
  \end{Large}
\end{center}
  \vspace{5mm}
\begin{description}
\item [Table I] Theoretical and experimental  $D^*$ and $B^*$
decay rates.
\end{description}
  \vspace{10mm}
\begin{description}
\item [Table II] Radiative decay widths of positive parity charmed mesons.
\end{description}

\newpage
\begin{table}
\begin{center}
\begin{tabular}{l c c }
 & {\bf Table I} &  \\ & & \\
 \hline \hline
Decay rate/ BR & theory & experiment \\ \hline
$\Gamma(D^{*+})$ & $46.1 \pm 14.2 \hskip 3 pt KeV$ &  $<131 \hskip 3 pt KeV \;$
\cite{ACCMOR}
\\ \hline
$ BR(D^{*+} \to D^+ \pi^0)$ & $31.2 \pm 17.4 \%$ & $30.8 \pm 0.4 \pm 0.8$ \\
\hline
$BR(D^{*+} \to D^0 \pi^+)$ & $67.7 \pm 34.2 \%$ & $68.1 \pm 1.0 \pm 1.3$ \\
\hline
$BR(D^{*+} \to D^+ \gamma)$ & $1.1 \pm 0.9 \%$ & $1.1 \pm 1.4 \pm 1.6$ \\
\hline \\ \hline
$\Gamma(D^{*0})$ & $36.7 \pm 9.7 \hskip 3 pt KeV$ &  \\ \hline
$ BR(D^{*0} \to D^0 \pi^0)$ & $56.4 \pm 27.1 \%$ & $63.6 \pm 2.3 \pm 3.3$ \\
\hline
$BR(D^{*0} \to D^0 \gamma)$ & $43.6 \pm 17.8 \%$ & $36.4 \pm 2.3 \pm 3.3$ \\
\hline  \\ \hline
$\Gamma(D^*_s)=
\Gamma(D^*_s \to D_s \gamma)$ & $(0.24 \pm 0.24) \hskip 3  pt KeV$ &  \\ \hline
\\ \hline
$\Gamma(B^{*+})=\Gamma(B^{*+} \to B^+ \gamma)$ & $(0.22 \pm 0.09)
\hskip 3  pt KeV$ &  \\ \hline \\  \hline
$\Gamma(B^{*0})=\Gamma(B^{*0} \to B^0 \gamma)$ & $(0.075 \pm 0.027)
\hskip 3  pt KeV$ &  \\ \hline \hline

\end{tabular}
\end{center}
\end{table}
\vspace{20 mm}
\vspace{20 mm}

\begin{table}
\begin{center}
\begin{tabular}{l c c}
 & {\bf Table II} &  \\ & & \\
 \hline \hline
Decay mode &  & width (KeV) \\ \hline
$ \Gamma (D^{\prime 0}_1 \to D^{*0} \gamma)$ & & $93 \pm 44 \hskip 5pt $ \\
\hline
$ \Gamma (D^{\prime 0 }_1 \to D^{0} \gamma)$ & & $14 \pm 6 \hskip 5pt $ \\
\hline
$ \Gamma (D_0^0 \to D^{*0} \gamma)$ & & $115 \pm 54 \hskip 5pt$ \\
\hline
$ \Gamma (D^{\prime +}_1 \to D^{*+} \gamma)$ & & $<  \hskip 5pt 2.3  $ \\
\hline
$ \Gamma (D^{\prime +}_1 \to D^{+} \gamma)$ & & $< \hskip 5pt 3.3 $ \\
\hline
$ \Gamma (D_0^+ \to D^{*+} \gamma)$ & & $< \hskip 5pt 2.8 $ \\
\hline \hline

\end{tabular}
\end{center}
\end{table}

\end{document}